# THE FUNDAMENTAL EQUATIONS OF POINT, FLUID AND WAVE DYNAMICS IN THE DE SITTER-FANTAPPIÉ-ARCIDIACONO PROJECTIVE RELATIVITY THEORY

Leonardo Chiatti
AUSL Medical Physics Laboratory
Via Enrico Fermi 15
01100 Viterbo (Italy)

**Summary**

A review is presented of the fundamental equations of point, perfect incompressible fluid and wave dynamics in the Fantappié-Arcidiacono theory of projective relativity, also known as "De Sitter relativity". Compared to the original works, some deductions have been simplified and the physical meaning of the equations has been analyzed in greater depth.

## 1. Introduction

This article proposes a modern introduction to point, fluid and wave dynamics, within the context of the theory of projective relativity developed by L. Fantappié (1901-1956) and later by G. Arcidiacono (1927-1998). We are actually dealing with two distinct theories: the theory of projective special relativity (PSR) and the theory of projective general relativity (PGR). The former is a generalization of the ordinary theory of special relativity (SR), postulating the invariance of physical laws with respect to the De Sitter group rather than to the Poincaré group, which is a local limit of it [1,2,3,4,5,6,7]. The latter is the corresponding generalization of the ordinary theory of general relativity (GR) [8,9]. The relation between PGR and PSR is the same as that between GR and SR. This article will deal exclusively with PSR, which has been restated by various authors under the name of "De Sitter relativity"; it has been discussed in various recent works [10,11,12,13,14,15,16].
PSR coincides locally with SR and its only difference from it lies in the predictions relating to the observation of objects that are very distant in space or events that are very distant in time; thus, crucial experiments (or, rather, observations) capable of confuting or verifying PSR can only be carried out in a cosmological context.
In this article, the Fantappié-Arcidiacono transformations that generalize ordinary SR Poincaré transformations will not be derived; for these preliminary aspects, the reader is referred to other works [17,18,19,20,21]. After an introduction recalling the kinematics of PSR (Sect. 2, 3), the fundamental equations of point (Sect. 4), perfect incompressible fluid (Sect. 5, 7) and wave (Sect. 8) dynamics will be introduced. Compared to the original Italian-language works, various deductions have been simplified and some errors have been corrected; also, the physical meaning of equations has been discussed in greater depth. Some comments on the physical meaning of quantities in PSR (Sect. 6, 9) have also been added; indeed, this is a topic which can give rise to misunderstandings.

## 2. PSR metric

In PSR, five projective coordinates, $\underline{x}_0, \underline{x}_1, \underline{x}_2, \underline{x}_3, \underline{x}_5$, are used, which are linked to the physical coordinates $x_0, x_1, x_2, x_3, x_5$, by the relation:

$$x_i = (\underline{x}_i/\underline{x}_5)r \qquad i = 0, 1, 2, 3 . \tag{2.1}$$

From here on, we shall use the indices $i, j, k, l, m...$ for the values 0, 1, 2, 3 and the indices $a, b, A, B,...$ for the values 0, 1, 2, 3, 5; the Greek indices $\mu, \nu$ will be used when referring only to the spatial coordinates 1, 2, 3. The coordinate $x_0$ is $ict$, where $t$ is the chronological distance from an observer, $c$ is the maximum speed of propagation and $i^2 = -1$. The constant $r$, having the dimensions of one length, is the radius of the De Sitter Universe[1]; the coordinates $x_1, x_2, x_3$ are the usual spatial coordinates, having their origin in the observer.

Equation (2.1) does not fix the value of $\underline{x}_5$; the Weierstrass condition is assumed:

$$\underline{x}_a \underline{x}^a = r^2 . \tag{2.2}$$

Thus, if we pose:

$$A^2 = 1 + \alpha^2 - \gamma^2 = 1 + \alpha_i \alpha^i , \tag{2.3}$$

with $\alpha_i = x_i/r$, $\gamma = ct/r = t/t_0$, it follows from equation (2.1) that:

$$\underline{x}_i = x_i/A ; \quad \underline{x}_5 = r/A . \tag{2.4}$$

Equations (2.1), (2.4) allow a coordinate $x_5 = r$ to be introduced; obviously, this is not a physical coordinate in the proper sense of the term, because it is not used by the observer to coordinate events [which occur in the continuum $(x_0, x_1, x_2, x_3)$]. The introduction of this coordinate facilitates expression of the correlation between data measured by different observers on the PSR chronotope; it must therefore be viewed in the sense of the intrinsic geometry of this chronotope rather than, extrinsically, as a manifestation of its curvature in an "external" five-dimensional space.

The projective metric is:

$$ds^2 = d\underline{x}_a d\underline{x}^a . \tag{2.5}$$

We observe that $r\underline{x}_i = x_i \underline{x}_5$, a relation which, when differentiated, gives:

$$rd\underline{x}_i = x_i d\underline{x}_5 + \underline{x}_5 dx_i . \tag{2.6}$$

By substituting equation (2.6) into (2.5) we obtain:

---

[1] This radius is a new fundamental constant in addition to $c$ and its introduction deserves comment. The invariance, with respect to inertial transformations, of the maximum propagation speed $c$ can be assumed as a postulate (Einstein's approach), or can be explained by the contact action of the "aether" on rulers and on clocks, in a Newtonian Galileo-invariant context (Lorentz-Poincaré approach). However, in order to explain the appearance of $r$ with this second mechanism, one would have to assume a non-local action by the aether on rulers of cosmic size, and this makes this approach decidedly less credible than the Einstein group approach. Thus, by adopting a group approach, one can pose oneself the problem of determining the largest four-dimension global invariance group that admits the Poincaré group as a local limit. This group is in fact the De Sitter group [1,2]. From this point of view, therefore, the PSR is the more general formulation of special relativity and ordinary SR is its limit case $r \to \infty$.

$$r^2 ds^2 = (dx_i dx^i)\underline{x}_5{}^2 + (r^2 + x_i x^i) d\underline{x}_5{}^2 + 2(x_i dx^i)\underline{x}_5 d\underline{x}_5 ,$$

and since $\underline{x}_5 = r/A$ it follows that:

$$d\underline{x}_5 = \alpha_i \, dx^i / A^3 \tag{2.7}$$

$$A^4 ds^2 = A^2 \, (dx_i dx^i) - (\alpha_i \, dx^i)^2 . \tag{2.8}$$

Equation (2.8) expresses equation (2.5) in terms of the physical coordinates; it is the metric on the geodetic representation of the De Sitter chronotope (known as the "Castelnuovo chronotope", [22,23,24]). The fundamental tensor associated to this metric is

$$g_{ik} = (A^2 \, \delta_{ik} - \alpha_i \, \alpha_k)/A^4 , \tag{2.9}$$

to which corresponds the counter-variant tensor

$$g^{ik} = A^2 \, (\delta^{ik} + \alpha^i \, \alpha^k) , \tag{2.10}$$

as it can be verified that:

$$g_{is} \, g^{ks} = \delta_i{}^k . \tag{2.11}$$

With a tedious but elementary calculation one has:

$$g = \text{Det} \, (g_{ik}) = A^{-10} . \tag{2.12}$$

The projective D'Alembert operator is thus obtained by using the general formula of mathematical analysis:

$$\Box \, \varphi = g^{-1/2} \, \partial_i \, (g^{1/2} \, g^{ik} \, \partial_k \, \varphi) , \tag{2.13}$$

from which we have:

$$r^2 \, \Box \, \varphi = A^2 \, (r^2 \, \partial_k \partial^k + x^i x^k \, \partial_i \partial_k + 2 \, x^i \partial_i) \, \varphi . \tag{2.14}$$

For $r \to \infty$, $\Box \to \Box = \partial_i \partial^i$. Wave propagation is described, in PSR, by equations as $\Box \, \varphi = 0$; this subject will be addressed later from a different viewpoint.

## 3. Kinematics of the material point

Equation (2.2) represents the hyper-spherical surface of radius $r$ having its centre at the origin, in a 5-dimensional Euclidean space $\{(\underline{x}_0, \underline{x}_1, \underline{x}_2, \underline{x}_3, \underline{x}_5)\}$. Let us consider the 4-dimensional space tangent to this hyper-sphere in a point that coincides with the observer; the hyper-spherical surface can be represented on this space by means of a projection from the centre of the sphere (this is known as a "geodetic" projection). Equation (2.8) is thus the Beltrami metric, induced on this space by the projection. This space is called "Castelnuovo chronotope", and it is within it that the observer coordinates events.

Each translation of a material point on the Castelnuovo chronotope is the projection of its motion over the surface (2.2); in other words, each translation on the "physical" chronotope actually is, in the 5-dimensional projective space, a rotation around the origin. Thus, in PSR, translations are a particular class of rotations. This implies that the equation of motion of a material point, rather than assuming the customary Newtonian form $\mathbf{F} = d\mathbf{p}/dt$, assumes a form which generalizes the equation $\mathbf{L} = d\mathbf{M}/dt$ valid for rotational motion ($\mathbf{F}$ = force, $\mathbf{p}$ = impulse, $\mathbf{M}$ = angular momentum, $\mathbf{L}$ = torque). From equation (2.8), posing $ds = icd\tau$, we have:

$$A^4 d\tau^2 = \left[A^2(1-\beta^2) + (\alpha \times \beta - \gamma)^2\right] dt^2 \qquad (3.1)$$

where $\beta = (\beta_0, \beta_1, \beta_2, \beta_3)$, $\beta_\mu = dx_\mu/(cdt)$, $\beta_0 = i$. From the identity

$$(\alpha \times \beta)^2 + (\alpha \wedge \beta)^2 = \alpha^2 \beta^2$$

it thus follows that:

$$A^4 d\tau^2 = \left[(1-\beta^2) + (\alpha - \beta\gamma)^2 - (\alpha \wedge \beta)^2\right] dt^2 = \left[B^2 - (\alpha \wedge \beta)^2\right] dt^2, \qquad (3.2)$$

where $B^2 = 1 - \beta^2 + (\alpha - \beta\gamma)^2$. We thus obtain the expression of the proper time interval $d\tau$. [2] Thus, the projective velocity:

$$\underline{u}_A = d\underline{x}_A/d\tau \qquad (3.2a)$$

and the projective acceleration:

$$\underline{a}_A = d\underline{u}_A/d\tau \qquad (3.2b)$$

can be introduced.
From equation $c^2 d\tau^2 = - d\underline{x}_A d\underline{x}^A$ it therefore follows that

$$\underline{u}_A \underline{u}^A = -c^2. \qquad (3.3)$$

By deriving equations (2.2), (3.3) with respect to $\tau$ we obtain the relations:

$$\underline{x}_A \underline{u}^A = 0 \ ; \quad \underline{u}_A \underline{a}^A = 0 \ ; \quad \underline{x}_A \underline{a}^A = c^2. \qquad (3.4)$$

The projective impulse is defined as:

$$\underline{p}_A = m_0 \underline{u}_A = m_0 d\underline{x}_A/d\tau, \qquad (3.5)$$

where $m_0$ is the local rest mass (i.e. the mass measured by an observer who is at rest with respect to the body *and who occupies the same position as the body*). It follows that:

$$\underline{p}_A \underline{p}^A = - m_0 c^2. \qquad (3.6)$$

Let us introduce the physical impulse as:

$$p_i = m_0 u_i = m_0 dx_i/d\tau. \qquad (3.7)$$

---

[2] In the two-dimensional case $(x,t)$ we have, starting from equation (3.1), $A^4 d\tau^2 = B^2 dt^2$.

From equation $x_i = r \underline{x}_i/\underline{x}_5$ it immediately follows that:

$$u_i = dx_i/d\tau = r\,(\underline{x}_5\,\underline{u}_i - \underline{x}_i\,\underline{u}_5)/\underline{x}_5^{\,2} \qquad (3.8)$$

and therefore

$$p_i = r\,(\underline{x}_5\,\underline{p}_i - \underline{x}_i\,\underline{p}_5)/\underline{x}_5^{\,2}\;. \qquad (3.9)$$

The projective angular momentum is defined as:

$$M_{AB} = \underline{x}_A\,\underline{p}_B - \underline{x}_B\,\underline{p}_A\;. \qquad (3.10)$$

By deriving equations (2.4) with respect to proper time, we have:

$$\begin{cases} A^3\,\underline{u}_i = \left(A^2\,\delta_{ik} - \dfrac{x_i\,x_k}{r^2}\right) u^k \\[2em] A^3\,\underline{u}_5 = -\dfrac{u_i\,x^i}{r}\;. \end{cases} \qquad (3.11)$$

By inserting equations (3.11) in equation (3.10) we have:

$$M_{5i} = r\,p_i/A^2\;;\quad M_{ik} = m_{ik}/A^2\;, \qquad (3.12)$$

where $m_{ik} = x_i\,p_k - x_k\,p_i$ is the usual physical angular momentum. From equations (3.10), (3.4) we obtain:

$$M_{AB}\,M^{AB} = 2r^2\,\underline{p}_A\,\underline{p}^A\;. \qquad (3.13)$$

Equation (3.13) can be expanded, using equations (3.12), in the form:

$$-2m_0^2\,c^2\,A^4 = \dfrac{m_{ik}\,m^{ik}}{r^2} + 2\,p_i\,p^i \qquad (3.14a)$$

or:

$$E = \pm c\sqrt{p^2 + m_0^2\,c^2\,A^4 + \dfrac{m_{ik}\,m^{ik}}{2r^2}}\;. \qquad (3.14b)$$

From equations (3.19), (3.4), we also obtain:

$$M_{AB}\,\underline{u}^A\,\underline{x}^B = m_0\,c^2 r^2\;, \qquad (3.15)$$

while from equations (3.11) and from the equation $\underline{x}_A\,\underline{x}^A = r^2$ one obtains:

$$A^3 p_i = p_i - x^k m_{ik}/r^2 ,  \qquad (3.16)$$

which is a relation between the impulse and the angular momentum. Finally, the projective moment of inertia tensor is introduced:

$$I_{AB} = m_0 \, \underline{x}_A \, \underline{x}_B . \qquad (3.17)$$

At small distances from the observer, $\underline{x}_i \approx x_i$ and $\underline{x}_5 \approx r$ whereby, within this limit:

$$I_{ik} = m_0 \, x_i \, x_k ; \quad I_{i5} = m_0 \, x_i \, r ; \quad I_{55} = m_0 \, r^2 . \qquad (3.18)$$

In other words, the ordinary moment of inertia, the static moment and the mass of the body are combined in $I_{AB}$.

## 4. Dynamics of the material point

The projective torque tensor is defined as:

$$L_{AB} = \underline{x}_A \underline{f}_B - \underline{x}_B \underline{f}_A ; \qquad (4.1)$$

in this definition, $\underline{f}_A$ is the projective force vector. Based on what has been said in the previous section, the equation of motion is

$$\frac{dM_{AB}}{d\tau} = L_{AB} . \qquad (4.2)$$

The concept of "free material point" requires some attention. According to equation (4.2) this type of body is characterized by the condition $L_{AB} = 0$; now:

$$\frac{dM_{5i}}{d\tau} = L_{5i} = \underline{x}_5 \underline{f}_i - \underline{x}_i \underline{f}_5 ,$$

and for a free point we shall therefore have $L_{5i} = 0$. This condition in no way implies that $\underline{f}_i$ and $\underline{f}_5$ are simultaneously null, and indeed we shall see that they are not.
By virtue of equations (3.12), the condition $L_{AB} = 0$ becomes:

$$\frac{d}{d\tau}\left(\frac{p_i}{A^2}\right) = 0 ; \quad \frac{d}{d\tau}\left(\frac{m_{ik}}{A^2}\right) = 0 . \qquad (4.3)$$

On the other hand:

$$L_{AB} = \frac{dM_{AB}}{d\tau} = \frac{d}{d\tau}\left(\underline{x}_A \underline{p}_B - \underline{x}_B \underline{p}_A\right) = m_0 \left(\underline{x}_A \underline{a}_B - \underline{x}_B \underline{a}_A\right) ,$$

whereby $\underline{x}_A \underline{a}_B - \underline{x}_B \underline{a}_A = 0$. \qquad (4.4)

By multiplying both members of equation (4.4) by $\underline{u}^A$ and contracting on index $A$, we obtain the identity $0 = 0$; whereas, by multiplying them by $\underline{x}^B$ we obtain:

$$\underline{a}_A = H^2 \underline{x}_A, \qquad (4.5a)$$

where $H = 1/t_0 = r/c$. From equation (3.5) one thus has:

$$d\underline{p}_A/d\tau = m_0 H^2 \underline{x}_A. \qquad (4.5b)$$

Equation (4.5b) splits into the pair of relations:

$$d\underline{p}_i/d\tau = m_0 H^2 \underline{x}_i, \quad d\underline{p}_5/d\tau = m_0 H^2 \underline{x}_5.$$

By multiplying the first of these by $\underline{x}_5$, the second by $\underline{x}_i$ and subtracting one has:

$$\underline{x}_5 \frac{d\underline{p}_i}{d\tau} - \underline{x}_i \frac{d\underline{p}_5}{d\tau} = \frac{dM_{5i}}{d\tau} = m_0\left(\underline{x}_5 H^2 \underline{x}_i - \underline{x}_i H^2 \underline{x}_5\right) = 0,$$

and from the first of equations (3.12) one thus obtains the first of equations (4.3) again. Recalling equation (3.2), it takes the form ($V$ = velocity vector):

$$\frac{d}{dt}\left\{\frac{m_0 V}{\left[1-\beta^2 + (\alpha-\beta\gamma)^2 - (\alpha\wedge\beta)^2\right]^{1/2}}\right\} = 0. \qquad (4.6)$$

The solution of equation (4.6) is relatively easy in the two-dimensional case $(x, t)$; it becomes:

$$[1 + \alpha(\alpha-\beta\gamma)]\left(\frac{dV}{dt}\right) = 0. \qquad (4.7)$$

We have two solutions; one is constituted by uniform rectilinear motion $V$ = constant; the other is expressed by $\beta = (1 + \alpha^2)/(\alpha\gamma)$, which can easily be rewritten as:

$$\frac{dx}{dt} = \left(\frac{r}{t}\right)\left(\frac{x}{r} + \frac{r}{x}\right). \qquad (4.8)$$

Equation (4.8) is a differential equation with separable variables whose solution is:

$$x^2 - k^2 t^2 + r^2 = 0, \qquad (4.9)$$

where $k$ is an arbitrary constant. One immediately sees that in the observer's present ($t = 0$) one has $x = \pm ir$, an imaginary result. To avoid this singularity of kinematics, one must impose that these bodies not be simultaneous to any observer, but this is tantamount to saying that they are not physical. In other words, the solutions of equation (4.8) are not physically admissible and must be ruled out; with this exclusion, the only possible free motion remaining (in the two-dimensional case) is uniform rectilinear motion.

At this point, a digression is necessary. Let us consider equation (3.14b) again, which we rewrite in the form

$$p^2 - \frac{E^2}{c^2} + \frac{m_{ik}\, m^{ik}}{r^2} = -m_0^2\, c^2\, A^4 \ . \tag{4.10}$$

For a material point at rest, $p = 0$ and $m_{ik} = 0$, so that:

$$E = m_0\, c^2\, A^2 = m_0\, c^2 \left(1 + \frac{x^2}{r^2} - \frac{x_0^2}{r^2}\right) = m_0\, c^2 \left(\frac{r^2}{r^2} + \frac{x_\mu\, x^\mu}{r^2} - \frac{x_0^2}{r^2}\right) =$$

$$= m_0\, c^2 \left(\frac{x_5\, x^5}{r^2} + \frac{x_\mu\, x^\mu}{r^2} - \frac{x_0^2}{r^2}\right) = m_0\, c^2\, \frac{x_A\, x^A}{r^2} = m_0\, H^2\, x_A\, x^A \ ,$$

because $x_5 = r$. This, therefore, is the expression of rest energy in PSR. As regards *local* rest energy, it is expressed by:

$$m_0\, c^2 = \frac{E}{A^2} = m_0\, H^2\, \frac{x_A\, x^A}{A^2} = m_0\, H^2\, \underline{x}_A\, \underline{x}^A \ .$$

Let us therefore assume the following expression for the energy tensor of the free material point:

$$T_{AB} = m_0\, (\underline{u}_A\, \underline{u}_B - H^2\, \underline{x}_A\, \underline{x}_B) \ . \tag{4.11}$$

In this expression, the term $m_0 H^2\, \underline{x}_A\, \underline{x}_B$, whose spur is equal to local rest energy, is subtracted from the term $m_0\, \underline{u}_A\, \underline{u}_B$ which comes from the direct generalization of the similar SR expression. The term $m_0 H^2\, \underline{x}_A\, \underline{x}_B$ is null in the limit $r \to \infty$, in which SR is re-obtained.
To verify the validity of equation (4.11), let us define the projective force as[3]:

$$\underline{f}_A = \partial^B\, T_{AB} \ . \tag{4.12}$$

For a free material point we therefore have, considering that $\partial^A\, \underline{x}_A = 5$:

$$\underline{f}_A = -5 m_0 H^2 \underline{x}_A \tag{4.13}$$

i.e.

$$\underline{f}_i - \frac{x_i}{r}\, \underline{f}_5 = -5 m_0 H^2 \left(\underline{x}_i - \frac{x_i}{r}\, \underline{x}_5\right) = 0 \ , \tag{4.14}$$

because $x_i = r\, \underline{x}_i/\underline{x}_5$ . Now, the relation $M_{5i} = dL_{5i}/d\tau$ can be rewritten as:

$$\underline{x}_5\, \underline{f}_i - \underline{x}_i\, \underline{f}_5 = \frac{d}{d\tau}\left(\underline{x}_5\, \underline{p}_i - \underline{x}_i\, \underline{p}_5\right)$$

and, dividing it by $\underline{x}_5$:

---

[3] To avoid unduly complex notation from here on until Section 7 inclusive, we shall use the symbol $\partial_A$ to indicate the partial derivative with respect to the variable $\underline{x}_A$, rather than with respect to the variable $x_A$. In Section 8 the definition of projective derivation will be made explicit, and the related notation $\underline{\partial}_A$ will be introduced.

$$\underline{f}_i - \frac{x_i}{r}\underline{f}_5 = \frac{1}{\underline{x}_5}\left(\underline{\dot{x}}_5\,\underline{p}_i + \underline{x}_5\,\underline{\dot{p}}_i - \underline{\dot{x}}_i\,\underline{p}_5 - \underline{x}_i\,\underline{\dot{p}}_5\right) = \underline{\dot{p}}_i - \frac{x_i}{r}\underline{\dot{p}}_5 + \frac{1}{\underline{x}_5}\left(\underline{\dot{x}}_5\,\underline{p}_i - \underline{\dot{x}}_i\,\underline{p}_5\right).$$

On the other hand:

$$\underline{\dot{x}}_5\,\underline{p}_i - \underline{\dot{x}}_i\,\underline{p}_5 = \left(\frac{\underline{p}_5}{m_0}\right)\underline{p}_i - \underline{\dot{x}}_i\,\underline{p}_5 = \underline{p}_5\left(\frac{\underline{p}_i}{m_0}\right) - \underline{\dot{x}}_i\,\underline{p}_5 = \underline{p}_5\,\underline{\dot{x}}_i - \underline{\dot{x}}_i\,\underline{p}_5 = 0 \; ,$$

so that:

$$\underline{f}_i - \frac{x_i}{r}\underline{f}_5 = \underline{\dot{p}}_i - \frac{x_i}{r}\underline{\dot{p}}_5 \; .$$

Thus, from equation (4.14) we have, for a free material point:

$$\underline{\dot{p}}_i - \frac{x_i}{r}\underline{\dot{p}}_5 = 0 \; . \tag{4.15}$$

As a matter of fact, this equation is certainly valid because it can be derived from equation (4.5b), if we recall that $x_i = r\,\underline{x}_i/\underline{x}_5$. We can therefore conclude that equations (4.11), (4.12) are compatible with the dynamics of the free material point.
It is appropriate to point out that, also in the case of a free material point, we have $\underline{f}_A \neq 0$, as is clearly evidenced by equation (4.13). In PSR, it is the torque that is null in the free case, not the force; indeed, the spacetime translations are, in turn, rotations and therefore only rotations exist in reality. In free motion, the time variation of $\underline{p}_5$ cancels that of $\underline{p}_i$, as is evidenced in equation (4.15); in the two-dimensional case this implies uniform rectilinear motion.
We can obtain the same result by considering equation (4.11). The term $-m_0 H^2 \underline{x}_A \underline{x}_B$ depends on the coordinates: its divergence is therefore a force which, by acting on the point, determines its free motion. This force is precisely the left-hand of the equation (4.13).

One should stress that the conventional treatment of the De Sitter chronotope [25] does not make use of projective coordinates, and therefore $\underline{p}_5$ does not exist in that context. Furthermore, the motion equation is assumed to have the form $\boldsymbol{F} = d\boldsymbol{p}/dt$, rather than $\boldsymbol{L} = d\boldsymbol{M}/dt$. In the case of a free material point, this approach leads us to identify the force with expression (4.5b) which, for remote events that can be observed through their light and therefore placed on the observer's lightcone ($\alpha^2 = \gamma^2 \to A^2 = 1$ and $\underline{x}_A = x_A$), becomes $f_\mu = m_0 H^2 x_\mu$. In the conventional treatment, one has $H^2 = \lambda/3$, where $\lambda$ is the cosmological constant; thus, $f_\mu$ is nothing other than the "cosmological term". In other words, the disappearance of the "balancing" term $\underline{p}_5$ leads to a free motion which is no longer uniform but accelerated, and the force that must be introduced as the cause of this acceleration is the cosmological term.
It is possible to make free motion uniform again by suitably re-graduating clocks; this strategy leads to Milne's double time scale [26].

Having spoken of how the "cosmological term", non-existent in PSR, emerges in conventional theory, we ought now to speak of another important aspect of the De Sitter chronotope, to see how it is described in PSR: cosmic expansion.
In PSR, cosmic expansion derives from the transformations of coordinates which change one inertial system into another; these are Fantappié-Arcidiacono transformations, generalizations of

Lorentz-Poincaré transformations. It is therefore a *kinematic* and not a dynamic fact; this particular must be borne in mind.

The transformations relevant here are the time translations of parameter $T_0$; under one of these [17,18,19,20,21], the velocity $V$ of a body located in the event point $(x, t)$ of the unprimed reference frame becomes $V'$, where:

$$V' \sqrt{1-\gamma^2} = V\left(1 + \gamma \frac{t}{t_0}\right) - \gamma \frac{x}{t_0} \qquad (4.16)$$

and $\gamma = T_0/t_0$. If, in the unprimed reference frame, the body moves with uniform motion according to the law $x = Vt + x_0$ and $\gamma^2 \neq 1$, then

$$V' = \frac{V - \gamma \frac{x_0}{t_0}}{\sqrt{1-\gamma^2}}, \qquad (4.17)$$

which is a constant. Therefore, even in the primed reference frame the motion will be uniform rectilinear and its velocity will be $V'$. This is not a property peculiar to time translations but a common feature of all the transformations of the De Sitter-Fantappié-Arcidiacono group: they convert uniform rectilinear motions into uniform rectilinear motions. On the other hand this is nothing but a consequence of the covariance, with respect to that group, of equation (4.3) and its solutions.

From equation (4.17) it can be seen that for $\gamma \to \pm 1$, $V' \to \infty$ unless $V = \pm x_0/t_0$, a quantity which can assume a multiplicity of values, as the constant $x_0$ is arbitrary; in this case, equation (4.17) gives $V' = 0$. The first member of equation (4.16) thus tends to zero for $\gamma^2 \to 1$, and we obtain:

$$V = \frac{\gamma \frac{x}{t_0}}{1 + \gamma \frac{t}{t_0}} \; ; \qquad (4.18)$$

with $\gamma = \pm 1$ according to the sign of $T_0$. Equation (4.18) can be verified by directly substituting $x = Vt + x_0$ and $V = \gamma x_0/t_0$; the identity $V = V$ is obtained for every value of $\gamma$, thus also for $\gamma^2 \to 1$. In the $\gamma = +1$ case (past lightcone) we have $V = x_0/t_0$, where $x_0$ is the position of the body on the observer's simultaneity plane, and:

$$V = H(t)\, x \quad , \qquad (4.19)$$

where $H(t) = H/(1 + t/t_0)$, $-t_0 \leq t \leq 0$, $H = 1/t_0$. Equation (4.19) expresses the existence of a velocity field escaping from the observer, whose modulus increases with the distance from the latter; it is therefore a law of cosmic expansion. In the future lightcone ($\gamma = -1$), on the other hand, there is a cosmic contraction which is entirely symmetrical to this expansion, though it is not observable as it is not possible to receive signals from the future.

The not trivial fact is the compatibility between uniform free rectilinear motion and cosmic expansion. The field of velocity (4.19) has been derived from the request for non-divergence of the transformed velocity $V'$; it plays the role held by the "substratum" in Milne's kinematic relativity [27]. The dynamic equation (4.2) determines the local deviations from the "substratum" caused by the action of the forces. All this is unknown in ordinary special relativity.

The result obtained can be expressed by saying that the primed reference frame, or the system of bodies at rest with respect to it (for which $V' = 0$) exists if these bodies, in the unprimed reference frame, have velocities distributed in accordance with equation (4.19); i.e. if a cosmic expansion exists in this second reference frame. However, given that the choice of the primed reference frame is arbitrary, this result is equivalent to stating the existence of a class of observers who observe a cosmic expansion as described by equation (4.19); this class constitutes the "substratum". It is remarkable that the substratum should appears for merely kinematic (group) reasons, without any physical requirements such as the introduction of an aether might be.

**5. Dynamics of perfect incompressible fluids**

In SR the expression of the energy tensor of the perfect incompressible fluid is:

$$T_{ik} = (\mu + p/c^2)\, u_i\, u_k + p\delta_{ik} \;, \tag{5.1}$$

where $\mu$ and $p$ are the density and the pressure of the fluid, respectively, and $u_i$ is its quadrivelocity. The PSR generalization of equation (5.1) is obvious: one must substitute, in the limit $p \to 0$, the disgregated matter tensor $\mu\, u_i\, u_k$ with $\mu(\underline{u}_A\, \underline{u}_B - H^2\, \underline{x}_A\, \underline{x}_B)$. One thus obtains:

$$T_{AB} = (\mu + p/c^2)\, [\underline{u}_A\, \underline{u}_B - H^2\, \underline{x}_A\, \underline{x}_B] + p\delta_{AB} \;. \tag{5.2}$$

Let:

$$f^2 = \mu + p/c^2 \;, \tag{5.3}$$

and recalling that for a perfect incompressible fluid the equation of state[4]:

$$p = \mu c^2 \tag{5.4}$$

applies, equation (5.2) becomes:

$$T_{AB} = f^2\, \underline{u}_A\, \underline{u}_B + f^2 c^2\, [(1/2)\, \delta_{AB} - (1/r^2)\, \underline{x}_A\, \underline{x}_B] \;. \tag{5.5}$$

Thus, assuming that:

$$c_A = f\, \underline{u}_A \;, \tag{5.6}$$

the energy tensor becomes:

$$T_{AB} = c_A\, c_B - c^S\, c_S\, [\,(1/2)\, \delta_{AB} - (1/r^2)\, \underline{x}_A\, \underline{x}_B\,] \;. \tag{5.7}$$

It can be postulated [28,29] that this expression also remains valid in the more general case:

$$c_A = f\, \underline{u}_A + Q_A \;; \qquad Q_A\, \underline{x}^A = 0 \;; \qquad Q_A\, \underline{u}^A = 0 \;. \tag{5.8}$$

The relations obtained can be written in another form by introducing the generalized Eckart tensor:

---

[4] We recall that $\mu$ breaks down into a pure mass term $\mu_0$ and into a term dependent upon the specific internal energy $\varepsilon$ of the fluid, in accordance with the relation $\mu = \mu_0(1 + \varepsilon/c^2)$. The fluid is incompressible in the sense that in isothermal conditions $\mu_0$ is a constant; whereas $p$ obviously depends on the coordinates through $\varepsilon$. In these circumstances, the spacetime part of the fluid field $c_A$ is, in Einstein's $r \to \infty$ limit and in the absence of external forces, solenoidal [20, vol. II].

$$\eta_{AB} = \delta_{AB} + (1/c^2)\underline{u}_A\underline{u}_B - (1/r^2)\underline{x}_A\underline{x}_B \quad . \tag{5.9}$$

This tensor is symmetric, and in the proper reference all its components are locally null except the spatial ones $\eta_{\alpha\beta} = \delta_{\alpha\beta}$. It satisfies the conditions:

$$\eta_{AB}\underline{x}^A = 0 \; ; \quad \eta_{AB}\underline{u}^A = 0 \; ; \quad \Sigma_B \eta_{AB}\eta_{BC} = \eta_{AC} \quad . \tag{5.10}$$

Equation (5.2) becomes:

$$T_{AB} = \mu(\underline{u}_A\underline{u}_B - H^2\underline{x}_A\underline{x}_B) + p\eta_{AB} \quad , \tag{5.11}$$

while equation (5.7) becomes:

$$T_{AB} = c_A c_B - c^S c_S [\eta_{AB} - (1/2)\delta_{AB} - (1/c^2)\underline{u}_A\underline{u}_B] \quad , \tag{5.12}$$

an expression which keeps its form when the Einstein's limit $r \to \infty$ is performed.
From equations (5.10), (5.11) one has:

$$T_{AB}\underline{x}^A = -\mu c^2 \underline{x}_B \; ; \quad T_{AB}\underline{u}^A = -\mu c^2 \underline{u}_B \; ; \tag{5.13}$$

in other words, $\underline{x}_A$ and $\underline{u}_A$ are eigenvectors of $T_{AB}$ with eigenvalue $-\mu c^2$. Furthermore:

$$T_{AB}\underline{x}^A \underline{x}^B = T_{AB}\underline{u}^A \underline{u}^B = -\mu c^2 r^2 \quad . \tag{5.14}$$

The generalized Euler equations are obtained by equating to zero the divergence of equation (5.2); posing $f^2 = \mu + p/c^2 = m$ one has:

$$m\underline{a}_B + \underline{u}_B \partial_A(m\underline{u}^A) + \partial_B p - H^2 m \underline{x}_A \partial^A \underline{x}_B - H^2 \underline{x}_B \partial^A(m\underline{x}_A) = 0 \quad .$$

By multiplying this expression by $\underline{u}^B$ and $\underline{x}^B$, respectively, two continuity equations are obtained:

$$c^2 \partial_A(m\underline{u}^A) = dp/d\tau \tag{5.15a}$$

$$c^2 \partial_A(m\underline{x}^A) = dp/d\rho \tag{5.15b}$$

where $\tau$ is the curvilinear coordinate along the stream line, and $\rho$ is the spatial distance from the stream line. The expressions of the radial derivative ($d/d\rho = \underline{x}_A\partial^A$) and of the derivative along the stream lime ($d/d\tau = \underline{u}_A\partial^A$) have been taken into account. By substituting equations (5.15) into the principal equation, the generalized Euler equation is obtained:

$$m\underline{a}_B + (\underline{u}_B/c^2)dp/d\tau - (\underline{x}_B/r^2)dp/d\rho + \partial_B p = H^2 m \underline{x}_B \quad . \tag{5.16}$$

All the discussion conducted up to this point is valid for perfect fluids. In the case of viscous fluids, the term $-vV_{AB}$, where $v$ is the viscosity coefficient and $V_{AB}$ is the viscosity tensor obtained by directly generalizing the SR one [20], must be added to the second member of equation (5.12). One has:

$$2V_{AB} = \eta_{AR}\eta_{BS}(\partial^R c^S + \partial^S c^R) \quad . \tag{5.17}$$

## 6. Digression on the concept of temperature in PSR

Before explaining the fundamental equations of fluid with heat exchange in PSR, it is necessary to stop and discuss the concept of temperature in theories of relativity based on a global symmetry group. It is necessary to eliminate any ambiguity on the physical meaning of temperature as a quantity which will appear in those equations. The general problem of the meaning of physical quantities in PSR will be examined in Sect. 9.

Let us place ourselves in the context of ordinary SR, and let $T_0$ be the temperature of a gas measured by an observer at rest with respect to it; what is the temperature $T$ of this same gas measured by a second observer in uniform rectilinear motion at velocity $V$ with respect to the former? As is well known [30,31] there are, in SR, three distinct definitions of temperature which correspond to the three distinct laws of transformation:

$$T = T_0 \gamma \quad ; \quad T = T_0 \gamma^{-1} \quad ; \quad T = T_0 \quad , \tag{6.1}$$

where

$$\beta = V/c \quad ; \quad \gamma = 1/\sqrt{1-\beta^2} \quad .$$

The extension of these laws to the PSR domain is simple and obvious. Firstly, $T_0$ is the temperature measured by an observer who not only is at rest with respect to the gas, but is also located in the same spacetime region occupied by it. The contraction parameter $\gamma = dt/d\tau$ is generalized by the corresponding PSR quantity:

$$\Gamma = \frac{1 + \alpha^2 - \gamma^2}{\sqrt{1 - \beta^2 + (\alpha - \beta\gamma)^2 - (\alpha \wedge \beta)^2}} \quad , \tag{6.2}$$

where $\beta = V/c$, $\alpha = d/r$, $\gamma = t/t_0$. Here $d$ and $t$ are the parameters of the spacetime translation which transports the first observer into the second.

The temperature $T$ measured by the second observer is therefore, in accordance with the three distinct definitions:

$$T = T_0 \Gamma \quad ; \quad T = T_0 \Gamma^{-1} \quad ; \quad T = T_0 \quad . \tag{6.3}$$

It must be borne in mind [32] that the first "observer" is actually a thermometer, which must be in thermal equilibrium with the gas. Thus, it must be *at rest* with respect to the gas and *immersed* in it; the reading of this thermometer is therefore $T_0$. Even if the thermometer is read by an observer in motion with respect to it, or placed at cosmological distances from it, the result of the reading will always be $T_0$. Thus, if $T$ is understood as a "thermometer reading", one must necessarily have $T = T_0$. This supports the third definition (local proper temperature) and we shall use this one from now on.

# 7. Dynamics of perfect incompressible fluids with heat exchange

Arcidiacono studied, both in SR and in PSR, the case of a perfect incompressible fluid (described only by a single index $f$) subject to heat exchanges. He postulated the relation [28,29]:

$$c_A = f \underline{u}_A + Q_A \qquad (7.1)$$

with $Q_A \neq 0$, so that the hydrodynamic field $c_A$ is no longer parallel to the fluid stream $\underline{u}_A$. Precisely:

$$Q_A = q_A / f c^2 , \qquad (7.2)$$

where $q_A$ is the so called "thermal vector"; it satisfies the two conditions:

$$q_A \underline{x}^A = 0 , \qquad q_A \underline{u}^A = 0 . \qquad (7.3)$$

The thermal vector is linked to the absolute temperature $T$, defined in accordance with the previous section, by the generalized Fourier equation:

$$q_A = -\chi \eta_{AB} \partial^B T = -\chi \left[ \partial_A T + \frac{u_A}{c^2} \frac{dT}{d\tau} - \frac{x_A}{r^2} \frac{dT}{d\rho} \right] . \qquad (7.4)$$

In this equation, $\chi$ is the thermal conductivity coefficient, which we shall assume to be constant. By substituting equation (7.1) into equation (5.7), the energy tensor is obtained:

$$T_{AB} = f^2 \underline{u}_A \underline{u}_B + \frac{1}{c^2}(\underline{u}_A q_B + \underline{u}_B q_A) + \frac{q_A q_B}{f^2 c^4} + \left( f^2 c^2 - \frac{q^2}{f^2 c^4} \right)\left( \frac{\delta_{AB}}{2} - \frac{x_A x_B}{r^2} \right) , \qquad (7.5)$$

in which it has been posed $q^2 = q_A q^A$. We note that in the non-thermal case ($q_A = 0$) one has $c_A c^A = (f \underline{u}_A)(f \underline{u}^A) = -f^2 c^2 = -(\mu c^2 + p) = -2p$. Assuming the validity of the normalization $c_A c^A = -2p$ for $q_A \neq 0$, as well, one has:

$$f^2 c^2 - \frac{q^2}{f^2 c^4} = 2p . \qquad (7.6)$$

Recalling that $f^2 = m = \mu + p/c^2$ one can eliminate $f^2$ from equation (7.6), obtaining:

$$p^2 = \mu^2 c^4 - q^2/c^2 , \qquad (7.7)$$

a relation similar to that which applies in the relativistic hydrodynamics of SR.
Introducing the tensor $Q_{AB}$ by means of the expression:

$$c^2 Q_{AB} = \underline{u}_A q_B + \underline{u}_B q_A + q_A q_B / mc^2 , \qquad (7.8)$$

the equation (7.5) becomes:

$$T_{AB} = m \underline{u}_A \underline{u}_B - (2p/r^2) \underline{x}_A \underline{x}_B + p \delta_{AB} + Q_{AB} . \qquad (7.9)$$

By equating to zero the divergence of equation (7.9) one obtains:

$$m\underline{a}_B + \underline{u}_B \partial^A(m\underline{u}_A) + \partial_B p - \underline{x}_B \partial^A\left[(2p/r^2)\underline{x}_A\right] - (2p/r^2)\underline{x}_B + \partial^A Q_{AB} = 0 \ . \quad (7.10)$$

By multiplying this expression by $\underline{u}^B$, and bearing in mind that:

$$\underline{u}_B \underline{a}^B = 0 \ ; \quad \underline{u}_B \underline{u}^B = -c^2 \ ; \quad \underline{u}_B \partial^B p = dp/d\tau \ ; \quad \underline{x}_B \underline{u}^B = 0 \ ;$$

the continuity equation is obtained:

$$c^2 \partial^A(m\underline{u}_A) = dp/d\tau + \underline{u}^B \partial^A Q_{AB} \ . \quad (7.11)$$

Whereas by multiplying equation (7.10) by $\underline{x}^B$ and recalling that:

$$\underline{x}^B \underline{a}_B = c^2 \ ; \quad \underline{x}^B \underline{u}_B = 0 \ ; \quad \underline{x}^B \partial_B p = dp/d\rho \ ;$$

$$\underline{x}_B \underline{x}^B = r^2 \ ; \quad \underline{x}_A \partial_A \underline{x}^B = \underline{x}_B \ ; \quad 2\underline{x}^B \partial^A \underline{x}_B = \partial^A r^2 = 0 \ ,$$

one obtains:

$$r^2 \partial^A(2p\underline{x}_A/r^2) = mc^2 + dp/d\rho + \underline{x}^B \partial^A Q_{AB} - 2p \ . \quad (7.12)$$

By substituting equations (7.11), (7.12) into equation (7.10) one obtains:

$$m\underline{a}_B + (\underline{u}_B/c^2)dp/d\tau - (\underline{x}_B/r^2)dp/d\rho + \partial_B p - \partial^A Q_{AB} = H^2 m\underline{x}_B \ . \quad (7.13)$$

This is the Euler equation proposed by Arcidiacono for perfect incompressible fluids with thermal exchange. When the thermal vector vanishes, this equation is reduced to equation (5.16).

## 8. D'Alembert equation

In Section 2 the D'Alembert projective operator, which rules free wave propagation, was introduced starting directly from the metric. In this section, we propose a different and instructive construction, starting from the projective derivatives [20, vol. II].
By differentiating the equation:

$$x_a = (\underline{x}_a/\underline{x}_5)r \quad (8.1)$$

one obtains:

$$\frac{d\underline{x}_a}{r} = -\frac{\underline{x}_a}{\underline{x}_5^2} d\underline{x}_5 + \frac{d\underline{x}_a}{\underline{x}_5} \ ,$$

i.e.:

$$\frac{\partial \underline{x}_a}{\partial \underline{x}_5} = -r\frac{\underline{x}_a}{\underline{x}_5^2} \ ; \quad \frac{\partial \underline{x}_s}{\partial \underline{x}_a} = \frac{r}{\underline{x}_5}\delta_{sa} \ . \quad (8.2)$$

Let us define the projective derivation with respect to index $a$ as:

$$\underline{\partial}_a \varphi = (\partial \varphi / \partial \underline{x}_a) = \sum_s \left(\frac{\partial \varphi}{\partial x_s}\right)\left(\frac{\partial x_s}{\partial \underline{x}_a}\right) \ .$$

For $a \ne 5$ one has:

$$\underline{\partial}_a \varphi = \sum_s \left(\frac{\partial \varphi}{\partial x_s}\right) \delta_{sa} \frac{r}{\underline{x}_5} = \left(\frac{\partial \varphi}{\partial x_a}\right) \frac{r}{\underline{x}_5} = A \partial_a \varphi \ , \tag{8.3}$$

where $A = r/\underline{x}_5$. For $a = 5$ one has:

$$\underline{\partial}_5 \varphi = \sum_s \left(\frac{\partial \varphi}{\partial x_s}\right)\left(\frac{\partial x_s}{\partial \underline{x}_5}\right) = \sum_s (\partial_s \varphi)\left(-r \frac{x_s}{\underline{x}_5^2}\right) = -A \sum_s (\partial_s \varphi)\left(\frac{x_s}{\underline{x}_5}\right) = -\frac{A}{r} \sum_s (\partial_s \varphi)\left(r \frac{x_s}{\underline{x}_5}\right) =$$

$$= -\frac{A}{r} x_s \partial^s \varphi \ . \tag{8.4}$$

In practice, the ordinary partial derivative with respect to the index $s = 5$ is the derivative with respect to the constant $x_5 = r$, and therefore it does not exist. The relations (8.3), (8.4) express the projective derivatives as a function of the ordinary ones. For $s = 0, 1, 2, 3$ one has:

$$\underline{\partial}_s \underline{\partial}^s \varphi = (A \partial_s)(A \partial^s) \varphi = A(\partial_s A)(\partial^s \varphi) + A^2 \partial_s \partial^s \varphi \ ,$$

and since $\partial_s A = x_s/(Ar^2)$,

$$\underline{\partial}_s \underline{\partial}^s \varphi = \frac{x_s}{r^2} \partial^s \varphi + A^2 \partial_s \partial^s \varphi \ .$$

Instead:

$$\underline{\partial}_5^2 \varphi = \left(-\frac{A}{r} x_l \partial^l\right)\left(-\frac{A}{r} x^m \partial_m\right) \varphi = \frac{A^2}{r^2} x_l \partial^l \varphi + \frac{A^2}{r^2} x_l x_m \partial^l \partial^m \varphi + \frac{x_l x^m x^l}{r^4} \partial_m \varphi \ ,$$

an expression in which the indices $l, m$ run along 0, 1, 2, 3. Since $x_l x^l = r^2(A^2 - 1)$, one has:

$$\underline{\partial}_5^2 \varphi = \frac{A^2}{r^2} x_l \partial^l \varphi + \frac{A^2}{r^2} x_l x_m \partial^l \partial^m \varphi + \frac{(A^2-1)}{r^2} x^m \partial_m \varphi \ .$$

At this point, the projective Dalembertian can be introduced:

$$\underline{\Box} \varphi = \underline{\partial}_a \underline{\partial}^a \varphi = (\underline{\partial}_s \underline{\partial}^s \varphi + \underline{\partial}_5 \underline{\partial}^5 \varphi) \ ; \quad s = 0, 1, 2, 3.$$

One immediately obtains:

$$\underline{\Box} \varphi = (A^2/r^2)(r^2 \partial_s \partial^s + x_l x_m \partial^l \partial^m + 2 x_s \partial^s) \varphi \ . \tag{8.5}$$

Equation (8.5) links the projective Dalembertian to the ordinary one $\partial_s \partial^s$. The D'Alembert wave equation thus takes the De Sitter-covariant form:

$$\square \varphi = 0, \tag{8.6}$$

and in this form has been extensively studied by Arcidiacono and Capelas de Oliveira [33,34].
It is to be noted that the components of the wave number vector $\underline{k}_a$ must be appropriately redefined in PSR. The plane wave $\exp(i\underline{k}_a\underline{x}^a)$ is a solution of equation (8.6) only if $\underline{k}_a\underline{k}^a = 0$. If $\underline{k}_0 = i\omega/c$ is defined as in SR, one must have that $\underline{k}_5 = \theta/r$, if one wants this component to disappear in the limit $r \to \infty$. The condition will then be satisfied if one lets:

$$\underline{k}_\alpha = n_\alpha \left[ (\omega/c)^2 - (\theta/r)^2 \right], \qquad \alpha = 1, 2, 3, \tag{8.7}$$

with $n_\alpha n^\alpha = 1$. The phase thus becomes $\underline{k}_a \underline{x}^a \to \mathbf{k} \cdot \mathbf{x} - \omega t + \theta$ for $r \to \infty$.

The static case, in which $\varphi$ not depends on time (i.e. on $x_0$), is very interesting. In this case, equation (8.6) becomes the generalized Poisson equation:

$$\underline{\Delta} \varphi = [\partial_\alpha \partial^\alpha + (x_\beta x_\gamma / r^2)\partial^\beta \partial^\gamma + (2 x_\alpha / r^2)\partial^\alpha] \varphi = 0, \tag{8.8}$$

where the Greek indices run along the ordinary spatial coordinates. In the case of a central field $\varphi = \varphi(\rho)$, $\rho = (x_\alpha x^\alpha)^{1/2}$, this equation admits of the solution [20,35]:

$$\varphi = - kY(\rho)/\rho, \tag{8.9a}$$

with

$$Y(\rho) = (1 + \rho^2/r^2)^{1/2} \{\cos [\operatorname{arctg}(\rho/r)] + \sin [\operatorname{arctg}(\rho/r)]\} . \tag{8.9b}$$

Note that for $r \to \infty$, $\varphi \to -k/\rho$, and this allows the constant $k$ to be physically identified. For example, in the case of the gravitational field it is clearly the mass of the attracting body, multiplied by the Newton gravitational constant.

## 9. The meaning of physical quantities in PSR

Let us consider two observers O and O' and let H(O|O') be the value of the physical quantity H in the place occupied by observer O but defined in the reference frame of observer O'. Let instead H(O'|O) be the value of the same quantity in the place occupied by observer O', defined in the reference frame of observer O. Let us then indicate with H(O|O) the value of H in the place occupied by observer O, as defined in the reference frame of O, and with H(O'|O') the value of H in the place occupied by observer O', defined in the reference frame of observer O'. The quantity H can be, for example, the gravitational or the magnetic field, the speed of light in the vacuum, etc.
That which observer O can actually measure, through an interaction, is H(O|O); similarly, observer O' can measure H(O'|O'). It is essential to understand that O cannot measure H(O'|O), nor can O' measure H(O|O'), because every measurement is an event and therefore is local. However, Fantappié-Arcidiacono transformations provide values of H(O|O') starting from, say, H(O|O); or the values of H(O'|O) starting from H(O'|O'). What, therefore, is the physical meaning of H(O|O'), H(O'|O) ?

One must bear in mind that the laws of propagation of physical phenomena formulated in the reference system of O give H(O'|O) as a function of H(O|O); the same laws, formulated in the system of O', connect H(O|O') to H(O'|O'). This is evident is one takes, as an example of the quantity H, a continuous field - magnetic, gravitational, etc. - though this restriction is not at all necessary. Thus, the "non measurable" quantity H(O|O') is related to the directly measurable quantity H(O'|O') through the laws of propagation; but H(O|O') can in turn be linked to the directly measurable quantity H(O|O) through Fantappié-Arcidiacono transformations. Thus, there actually is a link between two directly measurable quantities, namely H(O|O) and H(O'|O').

The difference between PSR and SR is that the parameter $r$ (or, which is the same, $t_0$) enters into both the passages which express this relation in PSR (law of propagation and transformation of the inertial reference frame), and therefore the causal link between distant events is affected by the global curvature of spacetime. Obviously, local interaction processes, i.e. those which involve energy exchanges over small distances compared to $r$ or over brief times compared to $t_0$, are not affected by the curvature. Therefore, physical quantities such as the dimension of bounded states (atoms, galaxies, etc...), the energy levels of bounded states, and so on, do not show any variation in PSR, whereas the link by means of signals between distant events does. For example, there will be a difference between the frequency of a light wave emitted by a galaxy, measured at the start, and the frequency of the same wave measured on its arrival in another galaxy. This is precisely what "red shift" consists of.

Though PGR has not yet been sufficiently investigated from this point of view, it is plausible that the topics discussed in this section can to a certain extent be relevant to it. The most important difference is that global reference frames associated with the observers O and O' no longer exist: the reference frames introduced by theory are now local. On the space tangent in O at the manifold X which generalizes the De Sitter chronotope, laws of propagation similar to those of PSR are still defined, and these still connect H(O'|O) to H(O|O). Yet, the relation between the quantity H(O'|O) thus introduced and the quantity H(O'|O'), defined in the origin of the space tangent in O' to X, is no longer expressed by global transformations such as the Fantappié-Arcidiacono ones. This relation is now expressed by the projective connection associated with the fundamental tensor of the metric which generalizes equation (2.5) [8,9,20,25].

In the practical use of PSR it is necessary accurately to define the suitable physical quantities of a problem, because the global curvature effects associated with spacetime translations (effects which do not exist in SR) can easily lead to paradoxes. Let us consider, for example, the case in which the quantity H is the spatial position $x$ of a material point in an isolated bounded system, and the concerned law of propagation is the equation of motion $x = x(t)$, solution of equation (4.2). This equation is valid in an inertial reference frame whose origin is in the observation pointevent O, and $t$ is the chronological distance from O. In this reference frame, the system to which the material point belongs is bounded and its centre of mass is assumed to be at rest; thus, one would be tempted to define the notion of "bounded system" by asserting that $|x| < R$, where $R$ is a constant of motion. As can easily be seen, this notion of "bounded system" is inconsistent in PSR, as it is incompatible with that of the inertial observer at rest with respect to the system. Such an observer evolves from event O to event O', which is the origin of a reference frame in which the event ($x$, $t$) is simultaneous with O'. From the general expression of coordinate transformations for time translations (one-dimensional case) one has [17,18,19,20,21]:

$$x' = \frac{x\sqrt{1-\gamma^2}}{1+\gamma\frac{t}{t_0}} \quad ; \quad t' = \frac{t+T_0}{1+\gamma\frac{t}{t_0}} \quad , \tag{9.1}$$

where $\gamma = T_0/t_0$. It follows, assuming the request for simultaneity $t' = 0$, that $T_0 = -t$ and:

$$x' = \frac{x}{\sqrt{1-\left(\dfrac{t}{t_0}\right)^2}} \ . \tag{9.2}$$

One clearly sees that in the translated reference frame the spatial position of the material point diverges for $t \rightarrow \pm t_0$, and therefore the notion of "bounded system" introduced above cannot be exported to the new reference frame. Physically, however, the observer is causally disconnected from events external to his lightcone, and therefore the divergence expressed by equation (9.2) does not have any consequences on how he sees the bounded system. The mistake consists in having introduced a notion of "bounded system" using a spatial position external to the observer's lightcone. This mistake can be remedied by introducing a different notion, which uses quantities *internal* to the lightcone. For example, one can say that the system is bounded in the sense that the travel of a ray of light from any of its parts to the observer have a duration not exceeding $R/c$, where $R$ is a constant of motion. As can easily be seen, this definition is invariant for time translations, i.e. it does not depend on the fact that the observer coincides instantaneously with O or O'.

One must however pay attention to the fact that while the duration of the light travel from one part of the system to the observer is invariant for time translations, the duration of the light travel between *a given* emission pointevent and *a given* observation pointevent such as O is instead changed by the action of transformations (9.1).

## 10. Concluding notes

Most works concerning PSR are available in Italian, and this fact has probably contributed to the limited dissemination of this theory among specialists. This work wishes to present a summary of the fundamental PSR equations that is comprehensible to a wider public and can lead on to more specialized studies. The fundamental dynamics equations of material point, perfect incompressible fluid and wave have been summarized and derived following more direct reasoning than can be found in the original texts. Some mistakes have been corrected or eliminated. For example, by generalizing Maxwell's equations in the De Sitter-invariant form on the Castelnuovo chronotope, a longitudinal component of the electromagnetic field appears, because of the finite value of $r$ [40], which satisfies equations similar to those of perfect fluid hydrodynamics [20]. Arcidiacono was convinced, on the basis of this purely formal analogy [36,37,38,39], that the longitudinal component *was* the hydrodynamic field! In this article, his "cosmic magnetohydrodynamic" model, based on these principles, has been completely ignored.

As can be seen from the discussion on material point dynamics (Sect. 4) and as confirmed by the right-hand members of equations (5.16), (7.13), an important difference with respect to SR is constituted by the dynamic effect of geodetic projection. In the conventional description of the De Sitter chronotope, this effect is, at least in part, recovered by introducing a "cosmological term" that does not exist in PSR. PSR thus becomes a useful model, at least to understand the possible *kinematic* origin of the cosmological term.

Again, on a kinematic basis, it is possible also to deduce a phenomenon of expansion of the Universe with a velocity field expressed by equation (4.19), which disappears in Einstein's limit $r \rightarrow \infty$. This field diverges at a chronological distance from the observer which is equal to $-t_0$, but this singularity - as has been discussed in other works [26,41] - cannot be identified with the big bang. Rather, it is a horizon dependent upon the observer.

Relation (3.14b) is very interesting. By applying it to the entire Universe, it would seem to suggest the possibility not only of inter-conversions of mass and energy but of angular momentum as well. This subject however is still hypothetical and is unexplored to date.

The projective effects do not affect interaction phenomena; these are still correctly described by SR, since they are local. For example, PSR cannot be taken as the basis to explain a cosmological variation of the fundamental constants. The projective effects, on the other hand, affect the propagation of signals between events that are distant in time and/or in space. For example, a discussion of travelling waves which are solutions of equation (8.6) shows [20] that they are subject to a Doppler effect that can be related to cosmic expansion. The frequency of the light wave emitted by a galaxy and measured in the reference frame of the emitting galaxy in the place of emission differs from the frequency of the same wave on its arrival in another galaxy, measured in the reference frame of the galaxy of arrival. This is what the cosmological "red shift" predicted by PSR consists of; its origin is entirely due to the Doppler effect and not to the variation of the distance scale (whereas in PGR there is a contribution deriving from this variation, [41]). This entire topic can be generalized to any quantity H, as illustrated in Sect. 9 and, with reference to temperature, in Sect. 6.

One hopes that this discussion can contribute to solving any doubts about the relation between "reality" and "appearance" in PSR, facilitating the approach to a theory which, in our opinion, deserves careful consideration.

**References**

1. Fantappié L.; *Rend. Accad. Lincei* XVII, fasc. 5 (1954)
2. Fantappié L.; *Collectanea Mathematica* XI, fasc. 2 (1959)
3. Arcidiacono G.; *Rend. Accad. Lincei* XX, fasc. 4 (1956)
4. Arcidiacono G.; *Collectanea Mathematica* X, 85-124 (1958)
5. Arcidiacono G.; *Collectanea Mathematica* XII, 3-32 (1960)
6. Arcidiacono G.; *Collectanea Mathematica* XIX, 51-72 (1968)
7. Arcidiacono G.; *Collectanea Mathematica* XX, 231-256 (1969)
8. Arcidiacono G.; *Collectanea Mathematica* XVI, 149-168 (1964)
9. Arcidiacono G.; *Collectanea Mathematica* XXXIV, 95-107 (1964)
10. Kerner H.E.; *Proc. Natl. Acad. Sci. USA* 73, 1418–1421 (1976)
11. Iovane G., Giordano P., Laserra E.; *Chaos Solitons Fractals* 22(5), 975–983 (2004) arXiv:math-ph/0405056v1
12. Aldrovandi R., Bertrán Almeida J.P., Pereira J.G.; arXiv:gr-qc/0606122v2 (2007)
13. Cacciatori S., Gorini V., Kamenshchik A.; *Ann. Der Physik* 17, 728-768 (2008)
14. Aldrovandi R., Pereira J.G.; *Found. Phys.* 39, 1-19 (2009)
15. Han-Ying G., Chao-Guang H., Zhan Xu, Bin Z.; arXiv:hep-th/043171v1 (2004)
16. Han-Ying G.; *Phys. Lett. B* 653(1), 88-94 (2007)
17. Arcidiacono G.; Projective relativity, Cosmology and Gravitation. Hadronic Press, Nonantum (USA), 1986
18. Arcidiacono G.; The theory of hyper-spherical universes. International Center for Comparison and Synthesis, Rome, 1987
19. Arcidiacono G.; *Hadron. J.* 16, 277–285 (1993)
20. Arcidiacono G.; La teoria degli universi, vol. I-II. Di Renzo, Rome, 2000 (in Italian)
21. Licata I.; *El. J. Theor. Phys.* 10, 211-224 (2006)
22. Castelnuovo G.; *Rend. Accad. Lincei* XII, 263 (1930)
23. Castelnuovo G.: *Scientia* 40, 409 (1931)
24. Castelnuovo G.; *Mon. Not. Roy. Astron. Soc.* 91, 829 (1931)

25. Pessa E.; *Collectanea Mathematica* XXIV, 151-174 (1973)
26. Licata I., Chiatti L.; to be appear in *Int. J. Theor. Phys.*(DOI 10.1007/s10773-008-9874-z, October 2008); arXiv:gr-qc/0808.1339 (2008)
27. Bondi H.; Cosmology. Cambridge University Press, Cambridge, 1961
28. Arcidiacono G.; *Collectanea Mathematica* XXIII, 105-128 (1972)
29. Arcidiacono G.; *Collectanea Mathematica* XXVI, 39-66 (1975)
30. Ott H.; *Zeits. Phys.* 175, 70 (1963)
31. Touschek B., Rossi G.; Meccanica Statistica. Boringhieri, Torino, 1970 (in Italian)
32. Hakim R., Mangeney A.; *Nuovo Cimento Lett.* I (9), 429-435 (1969)
33. Arcidiacono G., Capelas de Oliveira E.; *Hadron. J.* 14, 353 (1991)
34. Arcidiacono G., Capelas de Oliveira E.; *Hadron. J.* 14, 137 (1991)
35. Gomes D. ; O potential generalisado no Universo de De Sitter-Castelnuovo. Thesis, Campinas (Brazil), 1994 (in Portuguese)
36. Arcidiacono G.; *Rend. Accad. Lincei* XVIII, fasc. 4 (1955)
37. Arcidiacono G.; *Rend. Accad. Lincei* XVIII, fasc. 5 (1955)
38. Arcidiacono G.; *Rend. Accad. Lincei* XVIII, fasc. 6 (1955)
39. Arcidiacono G.; *Rend. Accad. Lincei* XX, fasc. 5 (1956)
40. Roman P., Aghassi J.J.; *J. Math. Phys.* 7, 1273 (1966)
41. Chiatti, L.: *El. J. Theor. Phys.* 15(4), 17–36 (2007). arXiv:physics/0702178